\documentclass{llncs}
\usepackage{pstricks}
\usepackage{pst-tree}
\usepackage{pst-node}

\usepackage{pst-tree}
\usepackage{pst-node}

\renewcommand{\figurename}{Fig.}

\renewcommand{\contentsname}{Contenido\\ \vspace{1cm} \hrule}
\usepackage[all]{xy}
\usepackage{amssymb}
\usepackage{amsmath}
\usepackage{latexsym}
\usepackage{floatflt}

\newcommand{\barray}{ %\left.
         \begin{array}{l}}
\newcommand{\earray}{\end{array}\  \ %\right] 
}
\newcommand{\vsp}{0.4pc}

\newcommand{\dplus}{\mathbin{\mbox{$+\!\!+$}}}

\newcommand{\member }{\ensuremath{\in\!}}

\newcommand{\comment}[1]{}

\newcommand{\la}{\ensuremath{\mathbin{\leftarrow}}}
\newcommand{\append}{\mbox{\it append}}

\newcommand{\br}{\mbox{$\,|\,$}}

\newlength{\headPlusArrowSpace}

\newlength{\headWithoutArrowSpace}

\newcommand {\nil} {[\,]}

\newcommand {\new} {\mbox{\it new}}

\newcommand{\bc}{\begin{enumerate}}
\newcommand{\ec}{\end{enumerate}}
\newcommand{\ba}{\begin{eqnarray}}
\newcommand{\ea}{\end{eqnarray}}
\newcommand{\bas}{\begin{eqnarray*}}
\newcommand{\eas}{\end{eqnarray*}}
\newcommand{\bp}{\begin{program}}
\newcommand{\ep}{\end{program}}

\usepackage{fancybox}

\usepackage{stmaryrd}
\renewcommand{\tablename}{Programa}
\newcommand{\selection}{\mbox{\it partition}}
\newcommand{\findmin}{\mbox{\it findmin}}
\newcommand{\divid}{\mbox{\it split}}
\newcommand{\perm}{\mbox{\it perm}}
\newcommand{\delete}{\mbox{\it delete}}
\newcommand{\newln}{\nonumber \\ & &\hspace{57pt}}

\newtheorem{program}{Program}
\newtheorem{clauses}{Clauses}
\newcommand{\leTerm}{\ensuremath{\le_{\tau}}}
\newcommand{\mysort}{\mbox{\it mysort}}
\newcommand{\sel}{\mbox{\it sel}}
\newcommand{\minlist}{\mbox{\it minlist}}

\newcommand{\union}{\mbox{\it union}}
\newcommand{\len}[1]{\ensuremath{\arrowvert #1 \arrowvert}}
\newcommand{\shuffle}{\mbox{\it shuffle}}
\newcommand{\eeqref}[1]{\mbox{{Clause}~\eqref{#1}}} %clref
\newcommand{\eeref}[1]{\mbox{{Clause}~\eqref{#1}}} %clref
\newcommand{\inssort}{\mbox{\it inssort}}
\renewcommand{\shuffle}{\mbox{\it shuffle}}
\newcommand{\select}{\mbox{\it filter}}
\newcommand{\selsort}{\mbox{\it selsort}}
\newcommand{\ord}{\mbox{\it ord}}
\newcommand{\co}{,}%{\hspace{0.05cm},\ \hspace{-0.05cm}}
\newcommand{\msort}{\mbox{\it msort}}
\newcommand{\comma}{\ensuremath{\wedge }}
\newcommand{\set}[1]{\ensuremath{\{#1\}}}
\newcommand{\rex}{\ensuremath{\lhd}}
\newcommand{\eqrex}{\ensuremath{\unlhd}}
\newcommand{\bce}{\begin{clauses}\begin{eqnarray}}
\newcommand{\ece}{\end{eqnarray}\end{clauses}}
\newcommand{\be}{\begin{eqnarray}}
\newcommand{\ee}{\end{eqnarray}}
\newcommand{\hsp}{\hspace{2cm}}
\renewcommand{\min}{\mbox{\it min}}
\newcommand{\noln}{\nonumber \\ 
& & \hsp \comma\ }

\newcommand{\ins}{\mbox{\it insert}}
\newcommand{\sortTS}{\mbox{\it sort\_TS}}
\newcommand{\sort}{\mbox{\it sort}}

\newcommand{\Fa}{\mbox{\it \bf F1}}
\newcommand{\Fb}{\mbox{\it \bf F2}}
\newtheorem{Property}{Property}

\title{%Sorting algorithms for free!\\
Deriving sorting algorithms via abductive logic program transformation
}
\author{Manuel Hern\'andez} %\and David A. Rosenblueth\inst{2}}

\institute{Instituto de Computaci\'on\\
Universidad Tecnol\'ogica de la Mixteca\\
C.P. 69000, Huajuapan de Le\'on, Oaxaca, M\'exico\\
\email{\tt manuelhg@mixteco.utm.mx}}
\begin{document}

\maketitle

\begin{abstract}
Logic program transformation by the unfold/fold method advocates the
writing of correct logic programs via the application of some rules to
a naive program. This work focuses on how to overcome
subgoal-introduction difficulties in synthesizing efficient sorting
algorithms from an naive sorting algorithm, through logic program
transformation and abductive reasoning.
\end{abstract}

\bibliographystyle{alpha}

\section{Introduction}

Logic program transformation (LPT) helps us to solve the following
problem: Given naive but inefficient logic program, find an efficient
version of this program.  The \emph{sorting problem} consists in
obtaining an ordered succession of comparable objects from an
unordered succession.  Because some human-guided transformations can
involve the adding of subgoals in the body of clauses, we take some
techniques from abductive logic programming (ALP) to justify the
selection these subgoals.  In this work we apply some transformational
techniques and abductive logic programming to a naive sorting
algorithm to derive some efficient sorting algorithms.

When we derive by transformational methods some of the sorting
algorithms, we note that such algorithms are consequence of specific
\emph{design decisions} implicit in the supporting definitions.
However, some other design decisions do not follow a strict deductive
analysis. It is required a complementary technique for synthesizing
some concrete sorting algorithms: some \emph{explanations} to be
entailed within a theory.  These explanations are carried out by
adding certain atoms to the body of some clauses, preparing this body
for some potential applications of the folding rule.  Because there
are many possible explanations, we have to justify how to obtain the
suitable explanations.

\subsection*{Structure of this work}
This work is organized as follows. Section 2 gives some preliminaries.
Section 3 presents some permutation and order-check algorithms. We exemplify
our transformations first deriving an $O(n^3)$ algorithm in Section 4;
next, $O(n^2)$ algorithms in Section 5 and 6, and finally an $O(n\log(n))$
algorithm in Section 7. Section 8 compares with related work,
and we finalize with some conclusions in Section 9.

\section{Preliminaries}
We assume a basic familiarity with logic programming.  Now, let $S$ be
a finite sequence of objects, where these objects are taken from an
set $A$, with $A$ having a complete order relation.  The \emph{sorting
  problem} consists in finding an ordered version sequence $T$ from an
unordered sequence $S$.  A naive algorithm for solving the sorting
problem relies on considering $T$ is a permutation of $S$.

In the \emph{insertion algorithm} we take $S$ and $T$ as lists, $T$ is
initially the empty list, and then we proceed as follows: We take the
first element of $S$, $a$, and we insert $a$ into $T$. Now we take the
second element of $S$, $b$, and we insert $b$ in the \emph{correct}
position within $T$, and so on, until list $S$ be empty.

In the \emph{selection algorithm} we select a minimal element of $S$,
$a$, we delete $a$ from $S$, obtaining $S'$ and we place $a$ into a
new list $T$.  Now we proceed to deal with $S'$ to find another
minimal element $b$ of $S'$, deleting from $S'$, and placing it into
$T$ \emph{after} $a$, and so on, until list $S$ be empty.

For the \emph{mergesort algorithm} we split $S$ into two lists, almost
of the same size, $S_1$ and $S_2$. Next, we sort $S_1$ and $S_2$, and
finally, we merge $S_1$ and $S_2$ into a new list $T$ by intermixing
their elements always placing them in the correct position.

Finally, the \emph{quicksort} algorithm proceeds as follows: Given the
list $S$, we select an arbitrary element $a$ in $S$.  Now we partition
$S$ into two sublists, $S_1$ and $S_2$, with $S_1$ consisting of those
elements of $S$ being less or equal to $a$, and $S_2$ consisting of
those elements of $S$ being greater than $a$. We apply recursively the
same quicksort algorithm to $S_1$ and $S_2$, obtaining $S_1'$ and
$S_2'$, respectively, and the final result is the concatenation of
$S_1'$, ${a}$, and $S_2'$.

It has been observed \cite{Darlington:sssa} that the insertion
algorithm is a particular case of the mergesort algorithm, and the
selection algorithm is a particular case of the quicksort
algorithm. Also, the mergesort and the quicksort algorithms are
representative elements of algorithms following the general strategy
of \emph{divide-and-conquer} \cite{Smith85}.

\subsection{Abductive Logic Programming}

    Abductive logic programming (ALP) can overcome certain limitations
    of logic programming with respect to higher level knowledge
    representation and reasoning tasks \cite{alp1993}; in our case,
    the higher level knowledge is that of \emph{algorithms designed by
      human beings}, and the reasoning tasks are those of
    \emph{formally deriving through logic program transformation some
      of these algorithms}.

    ALP is a suitable framework for declarative problem solving that
    complements LPT, because in LPT the introduction of subgoals is
    not easily justified as significant and useful program development
    step; at least, not at the same degree as the elimination of
    clauses via the subsumption rule \cite{PPR:epdufr} or the
    application of the unfolding rule and simplification.
    % :
    % The program $(a)$ can be obtained from
    % $(b)$ by abduction; from the program $(a)$ we obtain
    % program $(b)$ by applying it the unfolding rule. 

There are some intuitive interpretations to understand 
this landscape of reasoning. We have:
%\begin{enumerate}
%\item 
(i) Deduction: If we have \emph{axioms $A$} and an \mbox{inference
  rule $R$}, we want to
  draw some \emph{conclusions $C$}.
%\item 
(ii) Induction: If we have \emph{causes $A$} and some \emph{effects} $C$, we
  want to know an \emph{inductive rule $R$}.
%\item 
(iii) Abduction: If we have an \emph{inference rule $R$} and some
  \emph{observations $C$}, we want to know the \emph{causes $A$} of these
  observations.
%\end{enumerate}

\comment{
%Beautiful framework:
From 
A R B
we have some unknown variables:
A R ?\hence B (deductive B).
A ? B \hence R (inductive R).
? R B \hece A (abduction)
}

\section{Abductive Logic Programming}
%Theory presentation: axioms and an inference-rule.

We give now the general framework of \emph{abductive logic} (AL).  Given a
\emph{theory presentation} $T$ and a sentence or observation $G$, the
AL problem consists in finding a \emph{set of sentences} $\Delta$, the
\emph{abductive explanation of} $G$, such that:
 (1) $T\cup \Delta \models G$, and
(2) $T\cup \Delta$ is consistent.
In the framework of \emph{abductive logic programming} (ALP), 
the theory presentation $T$ is a logic program $P$ (augmented with the
axioms of the Clark equality theory) and the SLDNF-resolution rule.

This an instance of this framework applied to LPT:
%
%Explanations:
%
%* Basic
%* Minimal
%* 
%Suppose a problem $P$, solved through a naive algorithm expressed as a
%logic program $I$, and an algorithm ${\cal A}$, expressed as the logic
%program $A'$, that
%efficiently solves the problem $P$.
%Let us formulate the following scenary: 
%After formulating some logic programs $\{C_1,\ldots,C_k\}$ ($k\le 1$)  
%as byproducts of a sequence of transformations in LPT, we observe that
%to obtain a. 
%
% expressed as a logic program $P$,
%
Let ${\cal A}$ be an algorithm that solves efficiently the sorting
problem.  Suppose that ${\cal A}$ 
is expressed as a logic program
$P$. Having cons\-truc\-ted a finite set of logic programs
$S=\{C_1,\ldots,C_k\}$ ($k\ge 1$) by transformation
(where $C_1$ solves directly the sorting problem), we have to find 
which significant
atoms are necessary to introduce as subgoals in the body of some
clauses belonging to the logic program $C_k$ for deriving the logic
program $P$ or a close variant. 
Now, $G$ is identified as the algorithm ${\cal A}$, expressed
as a logic program $P$. The
theory presentation $T$ is the set of logic programs
$S$  together with some properties obtained by the
context of the problem that the algorithm ${\cal A}$ solves.
The set $\Delta$ is a set of atoms to be introduced. 
The new theory
$T\cup \Delta$ should be consistent (integrity constraint), but in our
case %, by taking a finite set of Horn clauses as a logic program,
 the problem is simplified to selecting a significant set of atoms
and preserving completeness, because we already have the following important 
property: \emph{goal introduction always
preserves correctness although completeness can be altered by
thinning}. 
 Part of our contribution consists in finding 
what atoms are significant to preserve completeness. % \cite{DK2002}}.
\emph{Deleting subgoals within the body of some clauses}, in contrast to
goal introduction, increases the LHM. Also, correctness is
also preserved by deleting subgoals, but we should notice the possible
superset so created. 
%\begin{figure}
%\caption{}
%\end{figure}

Let P be a logic program. Let C be a clause belonging to P,  $C: p \la
q_1\wedge q_2\wedge \ldots\wedge q_m$ .  If we add a new subgoal $q$ to the body of $C$,
we have the new clause $C': p \la q\wedge q_1\wedge q_2\wedge
\ldots\wedge q_m$, and a new
logic program $P'$ that differs from $P$ only by clause $C$ and $C'$.
By taking the fix-point operator, we can obtain the least Herbrand
model of $P$ (LHM($P$)).  Now, through the application of the subgoal
introduction already made, we can conclude that LHM$(P') \subseteq$
LHM$(P)$; if LHM$(P') \subset$
LHM$(P)$, we have thinned the set LHM($P$) and we loose completeness
but not correctness. 
%We can conclude that subgoal introduction is an
%example of \emph{nonmonotonic reasoning}. 
%In this case Changing a clause through subgoal 
%introduction produces a reduction of the LHM(P).
 %
%The preferred explanations are those 
%having a minimal number of atoms and even so 
%
Integrity constraints, for us, is to try maintaining the original LHM
(completeness) because correctness is for sure. Our integrity
constraints are intended to maintain invariant the original 
certainty of the body of clauses.
In the following
sections we will look at specific techniques and examples of the application of these
general steps. %answers to these questions.

\section{Goal introduction keys}

In this section we want to characterize the 
abductive atoms to be added to the body of clauses. 
This characterization should give us criteria to define a suitable
search space $S$ %\cal{A}$
to identify some atoms as good candidates to continue with a
%useful 
transformation process.
%
%
%% Having described a fundamental property of the subgoal introduction
%% rule, our next task is to know how to obtain certain sorting algorithms.
%% To be more specific: Given the clause $p\la q_1,\ldots,q_m$ as
%% ocurring within an intermmediate logic program based on the the
%% application of logic program transformation rules, we ask
%% which atom
%% $q$ should be added to the succession $q_1,\ldots,q_m$ that allows us
%% to apply a folding step to get a known sorting algorithm or to follow
%% the general structure of a sorting algorithm.
  
%% An answer is to substitute some subgoals within the clause by a new 
%% atom: this is known as the definition rule. This is an analytical
%% view of  introducing atoms by replacing them by other already existent.
%%  A synthetic view is to introduce atoms (without substitution of other
%%  already
%% existent) 
%% having as predicate those belonging to the previous programs.
%% It is also in preference that these atoms together with other 
%% allow to  fold with respect to previous definitions. The arguments of
%% such atoms should be instantiated enough to allow to coordinate the 
%% variables as a whole.

Goal introduction is, at first sight, a pessimistic rule:
instead of decreasing the number of resolution steps, we  
increase it. However, we will use the goal introduction rule 
 as an
intermediate step
for the application of the folding
rule. We call both steps in sequence, an application of
the goal introduction followed by an application of the folding rule that
takes advantage of the atom added,
 \emph{abductive folding}.
We want to look for the best explanation (optimality
and utility in a certain sense) of the
 subgoal to be introduced. Let us consider the following 
logic program:
% and the constraint satisfaction
%given for the following circumstances:
\begin{eqnarray}
& &  \mbox{(a) } p \la q\wedge t \ \ \  \mbox{(b) }  s \la q\wedge r
  \ \ \  
 \mbox{(c) } u \la  m\wedge t\\
& &  \mbox{(d) }r \la \ \ \  \mbox{(e) } q \la \ \ \ \mbox{(f) } t\la
\end{eqnarray}
We want to add an atom $Q$ to the body of clause (a) for a posterior
folding. This example shows that we can fold in (a) with respect
$q$ and $r$, if $Q=r$, 
or with respect to $m$ and $t$, if $Q=m$. 
In the first case, we have $p \la s\wedge t$; in the second case, we 
have $p\la q \wedge u$. Later we will see why we prefer to fold 
with respect to $s$.
%(We can discard $Q$ as being $true$: correctness and
%completeness are valid, but is not interesting.)
%
%Integrity constraints are important... but not for us. 
%Instead, we have to impose 
In the next, we identify some
\emph{desirable properties of abductive explanations.}
 These properties facilitate the 
systematic application of the unfold/fold method, and more
specifically, the application of the folding rule.

\paragraph{Subgoals missing for applying the folding rule.} 
Our first desirable property of abductive explanations via subgoal
introduction is intended to satisfy the application of a folding rule.
At this point, we see goal introduction as an auxiliary rule for the
folding rule.  

% Also, goal introduction is best... when we have
%followed a general plan of derivation based on selecting suitable
%initial definitions and unfolding enough to recognize structural
%foldings.

\paragraph{Subgoals preserving successful paths.} 
Even with a good characterization of candidate atoms to be introduced,
subgoal introduction is by no means deterministic.
In the previous example we explore two possibilities for adding 
atoms: If $Q$ is $r$, we have: $p \la s\wedge t$. If $Q$ is $m$
then we have $p\la q\wedge u$.  We prefer to use the definition of $s$
because the query $\la s$ is successful, in contrast to $\la u$. Our
second desirable property for this choice here is to preserve completeness.
 %See p. 11, mod-survey.
Similarly, we prefer explanations participating on the major number of conclusions
(relevancy).
%This   is with respect to the least Herbrand model.
%%In the clause $p \la q\wedge Q\wedge t$ 
%we name Q an atom (or literal) to be
%introduced
%%we want  to make this clause \emph{true}. 
%This is a second integrity
%constraint à la logic programming. 
%%So we prefer $Q=r$ instead of $Q=m$.

\paragraph{Variable's coordination.}
Our third desirable property  is enunciated as a request to coordinate the
occurrence of variables. This is also better illustrated through an
example, this time involving predicates having variables as arguments:
\begin{eqnarray}
& & p(X,Y) \la q(X)\wedge r(Y)\ \ \  q(X) \la s(X) \ \ \  r(Y) \la t(Y) \\
& & m(X,Y) \la q(X) \wedge l(Y) \ \ \ n(X,Y) \la o(X)\wedge r(Y)
\end{eqnarray}
We introduce $l(Y)$ to obtain 
$
p(X,Y) \la q(X)\wedge l(Y)\wedge r(Y) %\label{dosast}
$
so that $Y$ is now to satisfy $l/1$ and $r/1$ instead of only $r/1$. This is
like a filter for discarding %(thinning)
 some possible terms $Y$.
Folding, we have $p(X,Y) \la m(X,Y)\wedge r(Y)$.
%
%In case of 
%\begin{equation}
%p(X,Y) \la q(X) \wedge l(Y1)\wedge r(Y)\label{oneast}
%\end{equation}
%we can fold thus: $p(X,Y) \la m(X,Y1)\wedge  r(Y)$
%but we only reclaim the existence of $Y1$, without any link to
%variable $Y$.
%The problem here is whether we can give a meaning to recover $Y1$
%but in this case $X$ is constrained without a clever reason.
The point is: \emph{We do not constraint original variables 
unless they are linked to others.} 
%We prefer \eqref{dosast} to \eqref{oneast}. 
% Or in any case, we would have:
%\begin{equation}
%p(X,Y) \la  q(X)\wedge o(X)\wedge r(Y)
%\end{equation}
%and folding, we have $p(X,Y) \la q(X)\wedge n(X,Y)$ with some general
%notes by this time to $X$.

\paragraph{General constraints of ALP.} 
Other constraints are valid here as in abductive logic in general:
minimality ($o(X)$ is preferred to $o(X)\wedge o(X)$)
and most specific terms to link subgoals through variables:
 we prefer $o(X)$ instead of $o(g(X))$ or we prefer $o(X)$ to $o(a)$ ($a$ is a constant).

\paragraph{Occam's razor.}

Finally, we look for \emph{properties} obtained from the existent subgoals.
%the major information possible. 
Because goal introduction is so
demanding (because the big search space), we want to exploit the information
already provided by the existent atoms instead of introducing new ones
or, at least, to prepare the ground for introducing new ones.

%Knowledge assimilation is something simple: 
If we add atoms to the
body of clauses, these atoms should be characterized in some of the 
following types (see page 10, mod-survey.dvi):
\begin{enumerate}
\item The new information is already deducible from the current atoms; 
we can explicitly extract information from the existent atoms through 
semantic domain \emph{properties}.
\item  Some parts are subsumed: we can delete them.
\item The new subgoal leads to contradiction:  clause would be erased. 
We have to avoid having an implosion (LHM$(P)=\emptyset$).
So that we should be careful about collapsing the LHM. 
\item The new information cannot derived from the current atoms.
\end{enumerate}
Other cases are possible. For example, we can need new syntactical 
versions of terms.

%\section{The sorting problem and Permutation and order algorithms}

\section{Further Details about Abductive Folding}
In this section we elaborate on some details about the description and
the application of
abductive folding. 

Sometimes we need to discover the most direct way to explain a
conclusion.  For example, in the following program
(a) $p \la q$, (b) $q \la r$ and (c) $q \la s$,  on the one hand, there is an
explanation for $p$, namely, $q$. The conclusion $q$, on the other hand, 
has two possibles explanations: either $r$ or $s$. 

Now consider the following program:
%\begin{eqnarray}
%& & 
(a) $p\la q$ and (b) $q \la G$.
%%end{eqnarray}
Here we have an explanation for $p$, but we do not have any
explanation for $q$ (the meta-variable $G$ represents a possible
explanation).   By transitivity, really we do not have any
explanation for $p$ either. This transitivity can be made explicit as
follows: From $ p \la q$, and $q \la G$, %then 
by unfolding $q$ in $p \la q$ we get: $p \la G$. 
Now it is evident that we did not know how to conclude $p$.
We call $q$ a \emph{weak predicate}.
%\begin{Theo}
Weakness of predicates is made explicit through the unfolding rule. 
%\end{Theo}

Now suppose $q$ has two possible explanations: $r$ and $s$. 
If we know that $r$ is false, we discard $r$. If we know that $s$ is true, we
prefer $s$ to $r$.
%\begin{Theo}
To preserve the major possible completeness, we prefer true explanations
%\end{Theo}
because false (or nonexistent, in negation as failure) explanations 
lead us to failed paths.

%Método de las diferencias finitas (Paige).
%
%permutaciones y selección de pares
%
%universalidad y optimalidad de perm by merging (by selecting how to
%split) Algun ejemplo de perm y aplicación de este método.

%Unfolding processes can be under the carpet: 

%Direct explanations vs Non-direct explanations.

%Unfolding +weak predicates = weak predicates 

After finding some points where goal introduction would be possible,
we have to corroborate a sensible use of the subgoal within the body
of the clause. As we already seen, a first requirement is
that the subgoal contributes to apply the folding rule.  In
contrast to some proposal of abductive reasoning, here the predicate
of the atom used as subgoal can be the same than the head of the
clause. This is often discarded because we would incur in
\emph{petitio principii} (to beg the question, to call the
question), trying to explain an effect through the same effect.
However, when arguments of predicates are given, we can use the
concept of \emph{well-founded recursion}: We can explain an effect
of an object %, object composed of minor objects, 
by the effects of the smaller constituent objects. 
%A procedural explanation would imply
%the information obtained through an analysis of the flow of
%computation, as a third possible requirement, more
%machine-implementation oriented.

%or if well-founded recursion is
%found or if computation-flow is acceptablte (moded programs?).

Now we consider the calculus of the subgoals missing for applying 
the folding rule. We show on Fig. \ref{usualfolding} the 
usual folding (by using in this example only one clause).
\begin{figure}
\begin{eqnarray}
\begin{array}{l}
\mbox{Fold in the clause}\\
\ \ \ p \la q_1\wedge  q_2 \wedge \ldots \wedge q_n \wedge A\\
\mbox{wrt}\\
\ \ \  q_1 \wedge q_2 \wedge \ldots \wedge  q_n\\ 
\mbox{by using}\\
\ \ \  C:\ q \la q_1 \wedge q_2 \wedge \ldots \wedge  q_n\\ \hline
\mbox{gives: } p \la q \wedge A
\end{array}&\hspace{1cm} & 
\begin{array}{l}
\mbox{To fold in the clause}\\
\ \ \  B:\ p \la q_2 \wedge q_3\wedge \ldots \wedge q_n \wedge A\\
\mbox{by using}\\
\ \ \  C: \ q \la q_1 \wedge q_2 \wedge \ldots \wedge  q_n\\ %\hline
\mbox{we need to add} \\
\ \ \  q_1 \mbox{  with substitution } \theta  \\ \hline
\mbox{gives: }  p \la q \wedge A
\end{array} \nonumber \\[3pt]
\mbox{(a) Usual folding.}& & \mbox{(b) Abductive folding.} \nonumber 
\end{eqnarray}
\caption{Usual and abductive folding.\label{usualfolding}}
\end{figure}

Abductive folding requires at least two steps, see (b) on
Fig. \ref{usualfolding}.
%% \begin{figure}
%% \[
%% \begin{array}{l}
%% \mbox{To fold in the clause}\\
%% \ \ \  B:\ p \la q_2 \wedge q_3\wedge \ldots \wedge q_n \wedge A\\
%% \mbox{by using}\\
%% \ \ \  C: \ q \la q_1 \wedge q_2 \wedge \ldots \wedge  q_n\\ %\hline
%% \mbox{we need to add} \\
%% \ \ \  q_1 \mbox{  with substitution } \theta  \\ \hline
%% \mbox{gives: }  p \la q \wedge A
%% \end{array}
%% \]
%% \caption{Abductive folding. \label{abdutivefolding}}
%% \end{figure}
Other steps such as calculating \emph{plain complements} or
\emph{coordinating variables} are also important. We explain these concepts.

%
%Let $C: \ p \la q_1 \wedge q_2\ldots \wedge q_n$ be a clause. 
%Let $D:\ q\la  q_1\wedge \ldots \wedge q_{n-1}$ be another clause.
If we want to fold $p$ by using  $C$, we need to calculate
the \emph{plain complement} of folding some part of the body of $p$
by considering the body of $q$. The plain complement in the previous 
example is:
$
\{q_1,...,q_n\}\setminus \{q_2,\ldots,q_{n}\}=\{q_1\}
$
so that 
$\{q_1\}$ is the set of candidates to add to the body of B.
The plain complement of folding is given 
by trying to preserve the original
terms occurring in $\{q_2,\ldots,q_{n}\}$
in clause B and instantiating (by specialization) 

It is also necessary to speak about a \emph{strategy} of goal introduction.
In our case, we use a \emph{greedy} strategy to take benefit from
the occurrence of well-founded recursion: If we find a component of a possible folding
instantiated more closely to the base case, we introduce the plain
complement with a substitution $\theta$ for matching the folding clause.

%% Folding allows to recover the high-level predicates.

%% Let us consider the following formulation of perm:
%% \begin{eqnarray}
%% & &perm(Xs,Ys) \la split(Xs,Xs1,Xs2) \newln
%%    \perm(Xs1,Ys1) \noln
%%    \perm(Xs2,Ys2) \noln
%%    merge(Ys1,Ys2,Y)
%% \end{eqnarray}

Plain complements 
indicate us what atoms are missing to fold.
Well-founded search indicates us when is appropriate to fold 
to introduce well-founded recursivity.
%(We differ from mod-survey.ps.gz that we do not exclude conclusions
%from some clauses: we want this! (to create recursive calls).
%Also, we allow atoms having variables.)
Both techniques help us to implement an algorithmic strategy for
subgoal introduction, although this strategy is greedy and, in any
case, the strategy would require at least an approval by a human
being. The part of well-foundedness is taken from the partial order
$\leTerm$, when possible, defined over the Herbrand universe.
%(visto bueno humano).
%Residues: something is missing to carry out a folding... (heuristic).

%Folding monitoring (can we fold? can we fold? can we fold?
%Yes, almos we can fold... something is missing... can we introduce a
%subgoal? can we fold, can we fold? Not, it is not time... and so on.)

%Our set A of abducible predicates is every atom with every possible
%argument (gulp!).

To resume, we have described the following meta-algorithm:
\begin{enumerate}
\item Use the ``need-for-folding'' heuristics;
\item identify complements of atoms;
\item when atoms have arguments smaller than head-arguments, 
choose as candidate for folding;
\item fold.
\end{enumerate}

\comment{
X=(1+2)+3
Y=1+(1+2)
Associativity, conmutativity, term... what?

Let $(X_1,\ldots,X_n)$ be the variables in the body of clause (Gral). 
In order to preserve completeness, we have to ensure  that the 
domain D1xD2x..xDn is always the same.  When we have 
A\subset D1, thinning, etc. Universal cuantification, no thinning,
Theorem: Under subgoal introduction, thinning is possible.
Theorem: To preserve completeness we have to ensure thinning was not
ocurred.
Theorem: ...
}

%% Bayesian networks

%% a 
%% (a,b,0.3)
%% (a,c,0.2)
%% (a,d,0.5)
%% (d,e,0.2)
%% (d,f,0.8)
%% (c,e,0.7)
%% (c,g,0.3).
%% From a to g: 0.2*0.3, or not?

%% This new information requires support (new clauses with new
%% definitions).  The LHM is extended, although the original program is
%% even true.
%% }

%Instead of \models, deduction |-
%------------------------------------

To summarize, our abductive proposal is: First, considering a
\emph{good} algorithm ($O(n*\log(n))$, $O(n^2)$, $O(n^3)$) as is
already known in literature in a procedural way. Second, we transform
some clauses to obtain a structure already seen or known in this
algorithm.  Third, we add goals to the body of these clauses for
allowing to fold with respect to previous definitions. We apply the
folding rule and, finally, we check whether the selected sorting
algorithm has been obtained (when not, we can assess
whether the current version so obtained is good enough).

\section{Permutation and order-check algorithms}

To explain sorting algorithms from sorting by permutation 
adopt the following general guidelines: By using
a permutation algorithm based on merging, and goal introduction, 
we can explain mergesort and quicksort.
Similarly, by using a permutation algorithm based on insertion,
 and goal introduction, we
can explain insertion sort.  Finally, by using a permutation algorithm based on
selection, and goal introduction we can obtain the selection sort algorithm.
Permutation algorithms give us the skeleton of generation of some
sorting algorithms, and goal introduction allows to filter some
answers before using them, obtaining the concrete recursive calls.  

%\subsection{Some permutation algorithms in logic programmnig} 

We choose the least Herbrand model augmented with equations and
inequations $a<b$,
$a\leq b$ as constraints between terms $a$ and $b$ as the meaning of
logic programs.
Let us consider the sorting problem over the integers and the order
relation \emph{$a$ less than $b$} (written as $<$, from where we
derive by definition the notation of $\le,\ >$, and $\ge$).  The
following logic program solves the sorting problem correctly:
   %\begin{table}
%   \begin{program}[Naive algorithm] \label{sort:p1}
   \begin{equation}
   \sort(Ls_1\co Ls_2) \la \perm(Ls_1,Ls_2) \comma \ord(Ls_2)
   \end{equation}
 %  \end{program}
   %\caption{Un algoritmo}
   %\end{table}
   where $\perm(Ls_1,Ls_2)$ holds if
   the list $Ls_2$ is a permutation of the list  $Ls_1$, and
   $\ord(Ls_2)$ holds if the list $Ls_2$ is (non-decreasing)
   ordered by the %according to the
   relation  $\le$. 

 Under the SLD resolution rule, and given the definitions of $\perm$
 and $\ord$, the naive algorithm solves the sorting problem correctly,
 but if we suppose that the list $Ls$ has
 length $n$, the order of execution of this algorithm is $O(n!)$.
Our objective is to formulate algorithms more
efficient than the naive algorithm through LPT and ALP.
% It will result that the improved efficiency will
%depend on the $\perm/2$ and $\ord/1$ definitions.  

%(Sometimes we put some base cases over a same line for saving space.)

Let us
consider three permutation algorithms written as logic programs.  Each
permutation algorithm will determine the main \emph{structure} of a
distinct sorting algorithm (cf. \cite{Darlington:sssa}). Having this
structure, the next step will be to identify which atoms are necessary
to add to the body of some clauses (via suitable \emph{properties}
where these atoms appear) for a possible application of the folding
rule. Our idea is to apply the techniques previously described about 
plain complements and well-founded terms.
%(Search of well-foundness?).

  A set of axioms that defines a permutation of
  a list $Ls$ is:
% the
%following:
%\begin{table}
\begin{clauses}[Perm1, based on \ins]\label{sort:perm1}
\begin{eqnarray}
& & \perm1(\nil \co \nil ) \la \\
& & \perm1([A\br Ls_1]\co Ls_3) \la \perm1(Ls_1\co Ls_2)\comma
 \ins(A\co Ls_2\co Ls_3)\\[\vsp]
& & \ins(A\co Ls\co [A\br Ls]) \la \\
& & \ins(A\co [B\br Ls_1]\co [B\br Ls_2]) \la \ins(A\co Ls_1 \co Ls_2)
\end{eqnarray}
\end{clauses}
%\caption{Perm1, basadon en \ins}
%\end{table}

A second definition of a permutation algorithm is:
\begin{clauses}[Perm2, based on  \delete]
\begin{eqnarray}
& & \perm2(\nil \co \nil ) \la \\
& & \perm2(Ls\co [A\br Ls_1]) \la  \delete(A\co Ls\co Ls_2)\comma
 \perm2(Ls_2\co Ls_1)\\[\vsp]
& & \delete(A\co [A\br Ls]\co Ls) \la \\
& & \delete(A\co [B\br Ls_1]\co [B\br Ls_2]) \la  
    \delete(A\co Ls_1\co Ls_2) 
\end{eqnarray}
\end{clauses}

%(By the way, $insert(X,Ys,Zs) \la \delete(X,Zs,Ys)$,~\cite{SS:aop}, p.69.)
To show our third algorithm, first we present 
a generic definition of permutation, having a \emph{double} 
recursive structure:
\newcommand{\permG}{\mbox{\it permG}}

\begin{clauses}[PermG, based on \shuffle]
\begin{eqnarray}
& & \permG(\nil \co \nil ) \la \\
& & \permG([A]\co [A]) \la \\
& & \permG(Ls \co  Ls_5) \la 
  \union(Ls_1,Ls_2,Ls) \noln %--------------------------
  \permG(Ls_1\co Ls_3)  \noln
  \permG(Ls_2\co Ls_4) \noln 
  \shuffle(Ls_3\co Ls_4\co Ls_5)
\end{eqnarray}
\end{clauses}
where $\union(Ls_1,Ls_2,Ls)$ denotes a disjunct, nondeterministic
choi\-ce of the subsets $Ls_1, Ls_2$ of  $Ls$ such that
if $\append(Ls_1,Ls_2,Ls_3)$ and $\union(Ls_1,Ls_2,Ls)$ hold, then 
$\sort(Ls_3,Ms)$ and $\sort(Ls,Ms)$ for some sorting algorithm,
and 
the length of  $Ls$ is greater than or equal
to two. 
The \shuffle\ predicate is defined as
follows (\cite{Miranda86,Moller:acgsa}):
\begin{clauses}[Shuffle]
\be
& & \shuffle(\nil \co Ls\co Ls) \la \\
& & \shuffle(Ls\co \nil \co Ls) \la \\
& & \shuffle([A\br Ls_1]\co [B\br Ls_2]\co [A\br Ls_3]) \la %\newln
\shuffle(Ls_1\co [B\br Ls_2]\co Ls_3) \\
& & \shuffle([A\br Ls_1]\co [B\br Ls_2]\co [B\br Ls_3]) \la %\newln
\shuffle([A\br Ls_1]\co Ls_2\co Ls_3) 
\ee
\end{clauses} 
One way of translating  $\union(Ls_2,Ls_3,Ls_1)$ into a logic program 
is:
\bce
& & \divid(\nil \co \nil \co \nil ) \la \\
& & \divid([A]\co \nil \co [A])\la \\
& & \divid([A,B\br Ls_1]\co [A\br Ls_2]\co [B\br Ls_3])\la
\divid(Ls_1\co Ls_2 \co Ls_3)
\ece
The goal 
$\divid(Ls_1,Ls_2,Ls_3)$ divides the list  $Ls_1$ into two sublists, $Ls_2$ and
$Ls_3$, and $\len{Ls_2}=\len{Ls_3}$ or $\len{Ls_2}+1=\len{Ls_3}$,
where $\len{Ls}$ denotes the length of a list $Ls$.
 Now, we can formulate a concrete 
logical definition of our 
third permutation algorithm:
\begin{clauses}[Perm3, based on \divid\ and \shuffle]
\begin{eqnarray}
& & \perm3(\nil \co \nil ) \la \\
& & \perm3([A]\co [A]) \la \\
& & \perm3([A,B\br Ls_1]\co  Ls_6) \la %\newln
  \divid([A,B\br Ls_1]\co Ls_2\co Ls_3) \noln 
  \perm3(Ls_2\co Ls_4),\
  \perm3(Ls_3\co Ls_5)  \noln 
  \shuffle(Ls_4\co Ls_5\co Ls_6)
\end{eqnarray}
\end{clauses}

%12 [3]
%13 [2]
%31 [2]

\comment{RESERVED
The following algorithm allows to exchange two elements of a list,
giving us a new permutation algorithm:
\begin{clauses}[Perm4, based on exchanging elements]
\begin{eqnarray}
& & \perm4([],[]) \la \\
& & \perm4([A],[A]) \la \\
& & \perm4(Ls,Res) \la %\newln
	      \sel(T1,Ls,Ls_1) \noln 
	      \sel(T2,Ls_1,Ls_2) \noln 
	      \perm4(Ls_2,Ls_3)\noln
	      \append([T_1,T_2],Ls_3,Res)
\end{eqnarray}
\end{clauses}
RESERVED}

%\subsection{Order algorithms}

%To write specific definitions of \sort, 
We also need
some possible definitions of \ord. We give two definitions 
of \ord: \ord1 accesses consecutively elements of a list, 
and \ord2 delegates further comparisons to a predicate named \minlist.
%\ord1 is the traditional one and \ord2 is based on a
%predicate named \minlist.

\begin{clauses}[Ord1, linear]
\begin{eqnarray}
& & \ord1(\nil)\la \\
& & \ord1([A])\la \\
& & \ord1([A,B\br Ls])\la A \le B\comma \ord1([B\br Ls])
\end{eqnarray}
\end{clauses}
%Let us denote by $\set{Y}$ the set
%of elements of a list $Y$.
Another definition of \ord\ is the following:
\begin{clauses}[Ord2, subset]
\begin{eqnarray}
& & \ord2(\nil)\la \\
%& & \ord([A])\la \\
& & \ord2([A\br Ls])\la \minlist(A\co Ls)\comma \ord2(Ls)
%& & \minlist(-\infty\co \nil ) \la \\
%& & \minlist(A\co [A]) \la \nonumber \\
%& & \minlist(A\co [B\br L]) \la \nonumber \\
%& & \hsp \minlist(C\co L)\nonl
%       \min(B,C\co A)\\ [\vsp]
%& & \min(A,B,C) \la A<B\comma C=A \\
%& & \min(A,B,C) \la B<A\comma C=B 
\end{eqnarray}
\end{clauses}
where $\minlist(A\co Ls)$ holds if  $A$ is a lower bound
of all the
elements of the list $Ls$ ($A$ may or may not belong to the set of elements in
$Ls$), but it is only 
a check; $\minlist$ is not able to find this minimum.
\begin{clauses}
\begin{eqnarray}
& & \minlist(A,\nil) \la \\
& & \minlist(A,[B|Ls])\la A\le B\wedge \minlist(A,Ls)\label{minlist}
\end{eqnarray}
\end{clauses}
Theoretically, this predicate should be invertible, but the
inequality
in the body of Clause (\ref{minlist})
does not allow it; to adjust invertibility, we define a new predicate
to find the minimum of a list:
\begin{clauses}
\begin{eqnarray}
%& & \minlist(A,Ls) \la \findmin(B,Ls), A\le B. \\ [\vsp]
& & \findmin(-\infty\co \nil ) \la \\
& & \findmin(A\co [A]) \la \nonumber \\
& & \findmin(A\co [B\br Ls]) \la \findmin(C\co Ls)\wedge
       \min(B,C\co A)\\ [\vsp]
& & \min(A,B,C) \la A<B\comma C=A \\
& & \min(A,B,C) \la B\le A\comma C=B 
\end{eqnarray}
\end{clauses}

%% The order of a list can also be checked as follows: We divide the list $Ls$
%% into two sublists: $Ls_1$ and $Ls_2$, and then we proceed to say that
%% $Ls$ is ordered if $Ls_1$ is ordered, $Ls_2$ is ordered, and for all $X$
%% in the list $Ls_1$ and for all $Y$ in the list $Ls_2$, $X < Y$:
 
%% \begin{clauses}[Ord3, subsets]
%% \begin{eqnarray}
%% & & \ord3(Ls)\la 
%%          \append(Ls_1,Ls_2,Ls) \nonl
%%          \ord3(Ls_1),\ \ord3(Ls_2)             \nonl
%%           Ls_1 < Ls_2
%% \end{eqnarray}
%% \end{clauses}
%% %(something analogous would not be true for $\union/3$)
%% where $Ls_1 < Ls_2$ is defined as follows: for all $X$ in the list
%% $Ls_1$ and for all $Y$ in list $Ls_2$, $X\le Y$. 
%% %Other formulation: $Ls_1 < Ls_2$ if $last(Ls_1,Last)$ ,
%% %head(Ls_2,Head)$ hold then $Last < Head$.

Now, we need to formulate some \emph{properties} about
relationships between order-check and some other predicates 
that we use. We symbolize with $P\implies Q$ an implicative and
universally quantified logical formula between $P$ and $Q$.
%ocurring 
%whithin the body of some 
%clauses within our transformation steps.

%ocurring in the 
%definitions of permutation.
 
%about the relationship
%between $\append/3$ and $\ord/1$.
\begin{Property}[Append]
If  $\append(Ls_1,Ls_2,Ls_3)$ holds, % and 
\begin{equation}
\ord(Ls_3) 
%\mbox{ then }
\implies 
 \ord(Ls_1) \wedge \ord(Ls_2)
\end{equation}
\end{Property}

Let $C$ be a number and $Ls$ be a list of numbers. We use the
following notation: $C\rex Ls$ denotes
$C<D$, for every number such that $D\in \set{Ls}$, and $Ls\eqrex C$
denotes $D\le C$, for every $D\in \set{Ls}$.
%Now we use the following properties. First, we note that
\begin{Property}[Append and an element] \label{propApEl}
If $\append(Ls_1,[A|Ls_2],Ls)$ holds,
\begin{equation}
\ord(Ls)\ \ \vdash\ 
 \ord(Ls_1)\comma \ord(Ls_2) \comma Ls_1\rex A\comma
 A\eqrex Ls_2 
\end{equation}
\end{Property}
% $Ls_3= Ls_1 \dplus[A]\dplus Ls_2\comma \ord(Ls_3)$ implies where, if

%% \begin{Property}
%% If  $\append(Ls_1,Ls_2,Ls_3)$, 
%% \begin{equation}
%% \ord(Ls_1) \wedge \ord(Ls_2) \wedge Ls_1 < Ls_2 \implies \ord(Ls_3)
%% \end{equation}
%% \end{Property}

%Next, suppose $\subsequence(Ls_2,Ls_3)$ holds if $Ls_2$ is composed of some elements
%scattered along $Ls_3$.
%subq([],Ls).
%subq([A|Ls],[A|Ms]) :- subq(Ls,Ms).
%subq([A|Ls],[B|Ms]) :- subq([A|Ls],Ms).
%
%\begin{Property}[Subsequences\label{subsequences}]
%\begin{equation}
%\forall Ls_1,Ls_2: (
%\subsequence(Ls_1,Ls_2) \wedge \ord(Ls_2) \implies \ord(Ls_1) 
%\end{equation}
%\end{Property}

\begin{Property}[Insert]\label{insProp}
\begin{equation}
%\mbox{ If } 
\ins(A,Ls_1,Ls_2)\wedge\ord(Ls_2)
\implies 
%\mbox{ then } 
\ord(Ls_1)
\end{equation}
\end{Property}

\begin{Property}[Minlist]
If $\append([A],Ls_1,Ls_2)$ holds, 
\begin{equation}
\ord(Ls_2) \equiv
%\mbox{ is equivalent to }
 \minlist(A,Ls_1) \wedge \ord(Ls_1)
\end{equation}
%and 
%\begin{equation}
% \minlist(A,Ls_1) \wedge \ord(Ls_1) \implies \ord(Ls_2)
%\end{equation}
\end{Property}

\begin{Property}[Merging\label{shuffle}]
%{\em The merging lemma} \cite[p. 401]{Moller:acgsa},
%\[ (S\interleave T) \cap < = ((S\cap <) \interleave  T \cap
%<)) \cap <\  \]
%lemma translated in a predicative form as follows:
\begin{eqnarray*}
& & \shuffle(Ls_1\co Ls_2\co Ls_3)\wedge  \ord(Ls_3) \equiv 
%\mbox{ is equivalent to }
\\
& & \hspace{1.5cm}\ord(Ls_1)\wedge  \ord(Ls_2)\wedge 
\shuffle(Ls_1\co Ls_2\co Ls_3)\wedge \ord(Ls_3)
\end{eqnarray*}
\end{Property}

%% $12345 is ord if and only if 1 <2 and 2 <3 and 3<4 and 4<5$

%% $12345 is ord if minlist(1,2345) and minlist(2,345) and minlist(3,45)
%%                  and minlist(4,5),and minlist(5,[])-> true$

%% $12345 is ord if 123 is ord and 45 is ord and 3<4 (1);$

%% $12345 is ord if 123 is ord and 45 is ord X in 123 and Y in 45 implies X<=Y$ 

\section{Tamaki and Sato's sorting algorithm}

Let us begin with a reconstruction of a derivation 
of an $O(n^3)$ sorting algorithm. This derivation was given by Tamaki
and Sato in~\cite{TS:uftlp}, p.~135, from a definition of
{\it sort} given by \perm1 and \ord1. In the course of the derivation, 
an atom was added to the body of a clause as a subgoal. 
We will take this derivation as an instance of the previous general
results related to abductive folding: an application of the 
goal introduction rule followed by an application of the folding rule.

%Our reconstruction 
%will be based on the previous 
%attempts to explain how and why this 
%\emph{subgoal introduction} technique 
%works. 

%We begin by renaming the \sort predicate within the naive program.
To emphasize local transformational developments, 
we will apply a renaming of the \sort\ predicate at the beginning 
of each derivation. In this case we rename \sort\ to \sortTS.
\begin{program}[Naive program]
\begin{eqnarray}
\sortTS(Ls_1\co Ls_2) \la \perm1(Ls_1\co Ls_2)\comma \ord1(Ls_2) 
\label{sorting:naive}
\end{eqnarray}
\end{program}
Unfolding~\perm1\ in the body of~\eeref{sorting:naive} 
%(which
% is denoted by the subscript $U$),
 we obtain:
\begin{program}
\begin{eqnarray}
& & \sortTS(\nil \co\nil )\la %\ord(\nil )
\label{sorting:temp1}\\
& & \sortTS([A\br Ls_1]\co Ls_3) \la %\newln
         \fbox{\mbox{\ensuremath{\perm1(Ls_1\co Ls_2)\wedge \Fa}}} \noln 
         \ins(A\co Ls_2\co Ls_3) \wedge
         \ord1(Ls_3) \label{sort:1}
\end{eqnarray}
\end{program}
%$(A\equiv Q)  \wedge (Q\Rightarrow P)) \Rightarrow (A\equiv (Q \wedge P))$
%In the following, we silently unfold some subgoals like
%$\ord(\nil)$ in the body of~\eeref{sorting:temp1}.
%As it is pointed out by Tamaki and Sato, 

%% We are now guided by the ``need for folding'' step \cite{BD77:tsdrp}.
%% This is the part indicated by the metavariable $\Fa$.
%% In this case, the missing atom, indicated by $\Fa$, can be seen
%% like an ``explanation'' for the observation \emph{having a good sorting
%% algorithm}. To encompass human knowledge expertise, we need to
%% know which subgoals should be introduced within the body of the clause.
%% Suppose our criterion is applying the folding rule w.r.t. 
%% the original the definition of \sortTS, for having a recursive version 
%% of \sort.

By $Ls_1\leTerm[A\co Ls_1]$, reinforced with Property \ref{insProp}, 
%(which can be seen like an %******
%\emph{integrity
%constraint}),                      %
%\begin{eqnarray*}
%& & 
%###
%\wedge ord(Ls3)
%\end{eqnarray*}
%Then, 
we must add the subgoal $\ord1(Ls_2)$ to the body
of~\eeqref{sort:1} to proceed to apply abductive folding:
%\begin{clauses}
\begin{eqnarray}
& & \sortTS([A\br Ls_1]\co Ls_3) \la %\newln 
 \fbox{\mbox{\ensuremath{\perm1(Ls_1\co Ls_2)\wedge \ord1(Ls_2)}}}\noln
 \ins(A\co Ls_2\co Ls_3)\wedge \ord1(Ls_3)
\end{eqnarray}
%\end{clauses}
%(In the original paper \cite{TS:uftlp} this step is not justified at
%all.)

%The set of (not necessarily ground) abducible atoms \cite{alp1993} is either
%introduced by the definition rule \cite{PP:tlp1} or belongs to the set of the
%original atoms defined by the initial naive program.

% The query $Q$
%(within the ALP terminology) is satisfied when 
%the derived program preserves total correctness and (hopefully) we
%have a better efficiency than the observed in the original program.
%The lack of groundness guarantees that we can link the variables of 
%the subgoal introduced with those already existing in the clause. 

%As it was mentioned, the high-level knowledge guiding this goal
%introduction is the ``need for folding'' heuristics, but other
%heuristics like that of having a minor size of terms within recursive
%calls can be utilized. We can think of $\Delta$, the set of 
%abducible atoms, as consisting of 
%those (not necessarily
%ground) atoms that allow us to apply a folding step 
%w.r.t. an original or previous definition.

Folding  the subgoals 
 $\perm1(Ls_1\co Ls_2)$ and $\ord1(Ls_2)$ w.r.t.~$\sortTS$, 
%(subgoals marked with a $F$), 
we get a new program more efficient than the naive program:
\begin{program}[Tamaki \& Sato]
\label{sorting:ts}
\begin{eqnarray}
& & \sortTS(\nil \co \nil )\la \\
& & \sortTS([A\br Ls_1]\co Ls_3) \la %\newln
   \sortTS(Ls_1\co Ls_2)\noln 
   \ins(A\co Ls_2\co Ls_3)\noln
   \ord1(Ls_3)
\end{eqnarray}
\end{program}
We have obtained an $O(n^3)$ algorithm from an $O(n!)$ algorithm. The
moded program \cite{Apt:flptp97} described under the left-to-right
computation rule is: Given a list $[A\br Ls]$, we order $Ls$; next, we
insert $A$ in $Ls$ at some position; and, finally, we check whether the
resultant list is ordered.
%We would have a best
%performance if we could to test the order of the resultant
%list in terms of the {\em input list}, instead of doing it with
%respect to {\em output list}. In fact, we would be happy if
%we use only the input list as our unique list of work
%(because it is immediately known).
%Although 
%Tamaki and Sato assert to have, by derivation, an algorithm
%of order $O(n^2)$, based on the idea referred to another
%article, I have not found where is such a derivation.
%Anyway, 
Next, we will derive another algorithm of order
$O(n^2)$ from the naive program, the {\em insertion sort algorithm}. 
%Here, 7 sept 2002
% \begin{program}\label{sorting:insort}
% \begin{eqnarray}
% & & \inssort(\nil \co \nil )\la \\
% & & \inssort([A\br L]\co Ls_3) \la \newln
%        \inssort(L\co Z) \noln
%         \ins(A\co Z\co Ls_3) \noln 
% 	\ord(Ls_3) \noln
%         Z=Ls_1\dplus Ls_2\label{sorting:insortcl}
% \end{eqnarray}
% \end{program}

\section{The insertion sort algorithm}

Beginning now from Prog.~\eqref{sorting:ts},
let us rename the $\sortTS$ predicate to \inssort\ and apply a 
\emph{subgoal introduction} intended to take benefit from correctly
placing the element $A$.  
We divide the list $Zs$ into two
sublists $Ls_1$ and $Ls_2$ through 
the atom $\append(Ls_1,Ls_2,Zs)$.
%55
%%
Consider the following definition of insert:
\begin{eqnarray}
& & \ins(A,Zs,Ls) \la  
         \append(Ls1,Ls2,Zs) \wedge
         \append(Ls1,[A|Ls2],Ls)
\end{eqnarray}
%Now we introduce an atom $\append(Ls_1,Ls_2,Zs)$ (see Property 
%\eqref{propApEl}):
Now, we unfold insert w.r.t. this definition:

\begin{program}\label{sorting:insort}
\begin{eqnarray}
& & \inssort(\nil \co \nil )\la \\
& & \inssort([A\br Ls]\co Ls_3) \la %\newln
        \inssort(Ls\co Zs) \noln
        \append(Ls_1,Ls_2,Zs)\wedge
        \append(Ls_1,[A\br Ls_2],Ls_3)\noln
	\ord(Ls_3) \label{sorting:insortcl}
\end{eqnarray}
\end{program}

\begin{clauses}\label{clause:insort}
\begin{eqnarray}
%& & \inssort(\nil \co \nil )\la \\
& & \inssort([A\br Ls]\co Ls_3) \la %\newln
        \inssort(Ls\co Zs) \noln
        \append(Ls_1,Ls_2,Zs)\wedge
        \append(Ls_1,[A\br Ls_2],Ls_3)\noln
	\ord(Ls_3) \wedge
 \ord(Ls_1)\comma \ord(Ls_2) \wedge Ls_1\rex A\comma
 A\eqrex Ls_2 \label{sorting:insortcll}
\end{eqnarray}
\end{clauses}

%% Now we rewrite~\eeqref{sorting:insortcl}:
%% %\begin{clauses}
%% \begin{eqnarray}
%% %& & \inssort(\nil\co \nil )\la \\
%% & & \inssort([A\br Ls]\co Ls_3) \la %\newln
%%        \append(Ls_1\co Ls_2 \co Zs) \noln
%%        \inssort(Ls\co Zs) \noln
%%         \ins(A\co Zs\co Ls_3) \noln
%%         \ord(Ls_3)
%% \end{eqnarray}
%% %\end{clauses}
%If $A\in Zs$ then $\exists Ls_1Ls_2. Zs=Ls_1++[A]++Ls_2$.
%Suppose 

%The insertion of $A$ in $Zs$ given by the subgoal  
%$\ins(A\co Zs \co Ls_3)$%\hspace{\stretch{1}} 
%gives 
%$\append(Ls_1,[A|Ls_2],Ls_3)$, 
%and 
Because  $\ord(Ls_3)$ holds and the property that $ord(Ls)$ holds for every
sublist of $Ls_3$, we create a new definition 
that correctly places $A$ 
in the list $Ls_3$.
\begin{clauses}
\begin{eqnarray}
%& & \select(A\co\nil \co\nil \co\nil ) \la \\
& & \select(A\co\nil \co\nil \co\nil ) \la \\
& & \select(A\co [B\br Ls_1]\co [B\br Ls_2]\co Ls_3) \la 
       B\le A\noln
      \select(A\co Ls_1\co Ls_2\co Ls_3)\\
& & \select(A\co [B\br Ls_1]\co Ls_2\co [B\br Ls_3]) 
\la A< B\noln \select(A\co Ls_1\co Ls_2\co Ls_3)
\end{eqnarray}
\end{clauses} 
$\select(A\co Ls_3\co Ls_1\co Ls_2)$ holds if $Ls_3=Ls_1 \cup Ls_2$,
$Ls_1\rex A$, and $A\rex Ls_2$.

Replacing the subgoal given by~\eqref{sorting:insortcll}, we have an
\emph{insertion algorithm} $O(n^2)$ for correctly solving the sorting problem:
\begin{program} \label{insort:ins}
\begin{eqnarray}
& & \inssort(\nil \co \nil )\la \\
& & \inssort([A\br Ls_0]\co Ls_3) \la %\newln
       \inssort(Ls_0\co Zs) \noln 
       \select(A\co Zs \co Ls_1\co Ls_2) \noln
      \append(Ls_1\co [A|Ls_2]\co Ls_3)   
\end{eqnarray}
\end{program}
%% or, equivalently, expressing the
%%  Prog~\ref{insort:ins}  using~\append\:
%% \begin{program}[Insertion algorithm]
%% \begin{eqnarray}
%% & & \inssort(\nil \co \nil )\la \\
%% & & \inssort([A\br Ls]\co Ls_3) \la \newln 
%%        \inssort(Ls\co Zs)\noln
%%        \select(A\co Zs\co Ls_1\co Ls_2)  \noln
%% %       \append(Ls_1\co [A]\co Ys)  \noln
%      
%% \end{eqnarray}
%% \end{program}
%This insertion algorithm solves the sorting problem in order
%$O(n^2)$. 
Further minor
optimizations are possible 
(for example, we can use
difference lists instead of \append).

\section{Selection algorithm}

Let us see again the naive algorithm,
this time with the definition of \perm\ given by $\perm2$.
%\begin{program}
\begin{eqnarray}
& & \selsort(Ls\co Ls_1) \la \underline{\perm2(Ls\co Ls_1)}\comma  \ord2(Ls_1)
\label{sorting:selsort}
\end{eqnarray}
%\end{program}

 Unfolding~$\perm2$ in \eeqref{sorting:selsort}, we get
%\begin{figure}
\begin{program}
%\[
%\begin{array}{cc}
\begin{eqnarray}
& & \selsort(\nil\co \nil) \la \\
& & \selsort(Ls\co [A\br Ls_2]) \la %\newln 
        \delete(A\co Ls\co Ls_1) \noln
        \fbox{\mbox{\ensuremath{\perm2(Ls_1\co Ls_2)\wedge \Fa}}}\noln 
        \underline{\ord2([A\br Ls_2])}
\end{eqnarray} 
%&
%\begin{eqnarray}
%& & \selsort(L \co [A\br Ls_2]) \la \newln
%   \delete(A\co L\co Ls_1)\noln
%  \perm(Ls_1\co Ls_2)\noln  
%  A\rex Ls_2\noln
%  \ord(Ls_2)
%\end{eqnarray}
%\end{array}
%\]
\end{program} 
where we have made annotations about unfolding and a possible folding
point. %The numbers on the left indicate the order of 
%
%\end{figure}
%\setlength{\fboxrule}{1mm}
We have $\ord2([A\br Ls])$ if $\minlist(A,Ls)$  and
$\ord2(Ls)$. Replacing the subgoals, with $\Fa$ instantiated to
$\ord2(Ls_2)$,
we have:
%\begin{clauses}
\begin{eqnarray}
& & \selsort(Ls \co [A\br Ls_2]) \la %\newln
   \delete(A\co Ls\co Ls_1)\noln
  \fbox{\mbox{\ensuremath{\perm2(Ls_1\co Ls_2)\wedge \ord2(Ls_2)}}}\noln  
  \minlist(A,Ls_2)
\end{eqnarray}
%\end{clauses}
Now we fold the subgoals   $\perm2(Ls_1\co Ls_2)$  and  
$\ord2(Ls_2)$
w.r.t.~the original definition of~$\selsort$:
%\begin{clauses}
\begin{eqnarray}
%& & \selsort(\nil \co \nil) \la \\
& & \selsort(Ls\co [A\br Ls_2]) \la %\newln 
           \delete(A\co Ls \co Ls_1)\noln
           \selsort(Ls_1 \co Ls_2)\noln
           \minlist(A,Ls_2)
\end{eqnarray}
Now, $\minlist(A,Ls_2) \equiv \minlist(A,Ls_1)$; hence,
%See file selsort.pl
%\end{clauses}
%$delete_min(A,Ls,Ls_1) :-  findmin(A,Ls),delete(A,Ls,Ls1)$.
%Because $A\rex Ls_2$, and $\set{[A\br Ls_2]} = \set{L}$, $A\rex L$, 
%and then we obtain the final program:
\newcommand{\dm}{\mbox{\it delete\_min}}
\begin{eqnarray}
%& & \selsort(\nil \co \nil) \la \\
& & \selsort(Ls\co [A\br Ls_2]) \la %newln 
           \fbox{\fbox{\mbox{\ensuremath{\delete(A\co Ls \co Ls_1)\wedge 
           \minlist(A,Ls_1)}}}} \noln
           \selsort(Ls_1 \co Ls_2)
\end{eqnarray}
(We have surrounded a conjunction of atoms by a double box
anti\-ci\-pa\-ting the writing of a new definition and then the application
of a folding step w.r.t. this new definition.)

Making 
$\dm(A,Ls,Ls_1)\la  \delete(A,Ls,Ls_1)\comma minlist(A,Ls_1)$ 
or, 
\noindent  dually, 
$\dm(A,Ls,Ls_1) \la  \findmin(A,Ls)\comma \delete(A,Ls,Ls1)$
%
%and by applying some unfold/fold rules to this definition, we get:
%at:
%Unfolding ..
%delete_min(A,[A],[]).
%delete_min(A,[B|Ls],[B|Ls1]) :- delete(A,Ls,Ls1), minlist(A,[B|Ls1]).
%delete_min(A,[B|Ls],[B|Ls1]) :- delete(A,Ls,Ls1), A<B, minlist(A,Ls1).
%delete_min(A,[B|Ls],[B|Ls1]) :- A<B, delete(A,Ls,Ls1), minlist(A,Ls1).
%delete_min(A,[B|Ls],[B|Ls1]) :- A<B, delete_min(A,Ls,Ls1).
%\noindent 
we have:

\begin{program}[Selection algorithm]
\begin{eqnarray}
& & \selsort(\nil \co \nil ) \la \\
& & \selsort(Ls\co [A\br Ls_2]) \la  %\newln
           \dm(A\co Ls\co Ls_1)\noln
           \selsort(Ls_1\co Ls_2) \\[\vsp]
& & \mbox{plus definitions of \delete/3 and \findmin/2}\nonumber
\end{eqnarray}
\end{program}

\section{Mergesort algorithm}

To begin a new derivation, we rename \sort\ to \msort:
%\begin{program}
\begin{eqnarray}
& & \msort(Ls_1\co Ls_2) \la \underline{\perm3(Ls_1\co Ls_2)}\comma  \ord2(Ls_2)
\label{sorting:nai}
\end{eqnarray}
%\end{program}
%but now we proceed with the definition of \perm3.
We are now ready to unfold \perm3\ in the body of~\eeqref{sorting:nai}:
\begin{program}
\begin{eqnarray}
& & \msort(\nil \co \nil ) \la \\
& & \msort([A]\co [A])\la \\
& & \msort([A,B\br Ls_1]\co B) \la %\newln
        \divid([A,B\br Ls_1]\co Ls_2\co Ls_3) \noln
        \fbox{\mbox{\ensuremath{\perm3(Ls_2\co Ls_4)\comma \Fa}}}\noln 
        \fbox{\mbox{\ensuremath{\perm3(Ls_3\co Ls_5)\comma \Fb}}} \noln
        \shuffle(Ls_4\co Ls_5\co Ls_6)\wedge \ord2(Ls_6)
\end{eqnarray}
\end{program}
Now we add two subgoals, using Property \ref{shuffle}:

%\begin{clauses}
\begin{eqnarray}
%\bce
%& & \msort(\nil\co nil ) \la \\
%& & \msort([A]\co [A])\la \\
& & \msort([A,B\br Ls_1]\co Ls_6) \la %\newln
        \divid([A,B\br Ls_1] \co Ls_2\co Ls_3)  \noln
        \fbox{\mbox{\ensuremath{\perm3(Ls_2\co Ls_4)\wedge \ord2(Ls_4)}}}\noln
        \fbox{\mbox{\ensuremath{\perm3(Ls_3\co Ls_5)\wedge \ord2(Ls_5)}}}\noln
        \shuffle(Ls_4\co Ls_5\co Ls_6)\comma \ord2(Ls_6)
\end{eqnarray}
%\end{clauses}
%\ece
Now we can fold the subgoals $\perm3$ and $\ord2$ w.r.t.~the original definition of~\msort:
%\begin{clauses}
\begin{eqnarray}
%\bce
%& & \msort(\nil\co nil ) \la \\
%& & \msort([A]\co [A])\la \\
& & \msort([A,B\br Ls_1]\co Ls_6) \la %\newln
        \divid([A,B\br Ls_1] \co Ls_2\co Ls_3)  \noln
        \msort(Ls_2\co Ls_4) \noln
        \msort(Ls_3\co Ls_5) \noln
        \fbox{\fbox{\mbox{\ensuremath{\shuffle(Ls_4\co 
Ls_5\co Ls_6)\wedge \ord2(Ls_6)}}}}%_{\mbox{\tiny Def}}
\end{eqnarray}
%\end{clauses}
We write a new definition:
%\bce
%\begin{eqnarray}
$\new(Ls_1\co Ls_2\co Ls_3) \la  \shuffle(Ls_1\co Ls_2\co Ls_3)\wedge$
$ \ord2(Ls_3)$
%\end{eqnarray}
%\ece
and fold w.r.t.~this new definition:
%\begin{clauses}
\begin{eqnarray}
%\bce
%& & \msort(\nil\co nil ) \la \\
%& & \msort([A]\co [A])\la \\
& & \msort([A,B\br Ls_1]\co Ls_6) \la %\newln
        \divid([A,B\br Ls_1] \co Ls_2\co Ls_3)  \noln
        \msort(Ls_2\co Ls_4)\comma
        \msort(Ls_3\co Ls_5) \noln
        \new(Ls_4\co Ls_5\co Ls_6) \label{alreadyordered}
\end{eqnarray}
%\end{clauses}
From now on, our attention will be centered on \new.
Unfolding  \shuffle\  in the body of \new, we get the following clauses:
\begin{eqnarray} 
& & \new(\nil \co Ls,Ls)\la \ord2(Ls)\\
& & \new(Ls\co \nil \co Ls)\la \ord2(Ls)\\
& & \new([A\br Ls_1]\co [B\br Ls_2]\co [A\br Ls_3]) \la \newln
\hspace{-0.2cm}  \fbox{\mbox{\ensuremath{\shuffle(Ls_1\co [B\br Ls_2]\co Ls_3)\comma
\Fa}}}\comma
        \ord2([A\br Ls_3])\label{sorting:so1}\\
& & \new([A\br Ls_1]\co [B\br Ls_2]\co [B\br Ls_3]) \la \newln
\hspace{-0.2cm}	\fbox{\mbox{\ensuremath{\shuffle([A\br Ls_1] \co Ls_2\co
              Ls_3)\comma \Fb}}}\comma
        \ord2([B\br Ls_3])\label{sorting:so2}
\end{eqnarray}
We should replace the meta-variables $\Fa$ and $\Fb$ by concrete atoms
(or in general, literals) to allow the
application of the folding rule. This would give an explanation of 
having a $\new$ predicate self-contai\-ned, as is seen in
several versions of mergesort algorithms.  

\begin{eqnarray}
%\bce
& & \new([A\br Ls_1]\co [B\br Ls_2]\co [A\br Ls_3]) \la \newln
        \fbox{\mbox{\ensuremath{\shuffle(Ls_1\co [B\br Ls_2]\co Ls_3)
\comma
 \ord2(Ls_3)}}}  \noln
        \minlist(A,Ls_3) \\
& & \new([A\br Ls_1]\co [B\br Ls_2]\co [B\br Ls_3]) \la \newln
	\fbox{\mbox{\ensuremath{\shuffle([A\br Ls_1] \co Ls_2\co
              Ls_3)\comma \ord2(Ls_3)}}} \noln
        \minlist(B,Ls_3) 
%\ece
\end{eqnarray}

After folding, we obtain the following clauses:
\begin{eqnarray}
%\bce
& & \new([A\br Ls_1]\co [B\br Ls_2]\co [A\br Ls_3]) \la \newln
        \new(Ls_1\co [B\br Ls_2]\co Ls_3)\wedge 
        \minlist(A,Ls_3) \label{cp01}\\
& & \new([A\br Ls_1]\co [B\br Ls_2]\co [B\br Ls_3]) \la \newln
	\new([A\br Ls_1] \co Ls_2\co Ls_3)\wedge 
        \minlist(B,Ls_3) \label{cp02}
%\ece
\end{eqnarray}

Having folded, the next step is considering how to constrain the general 
behavior of $\minlist$ and its eventual elimination. 

%\begin{eqnarray}
%\bce
%& & \new([A\br Ls_1]\co [B\br Ls_2]\co [A\br Ls_3]) \la \newln
%        \new(Ls_1\co [B\br Ls_2]\co Ls_3),\ 
%        \minlist(A,Ls_3) \label{cp01}\\
%& & \new([A\br Ls_1]\co [B\br Ls_2]\co [B\br Ls_3]) \la \newln
%	\new([A\br Ls_1] \co Ls_2\co Ls_3),\ 
%        \minlist(B,Ls_3) \label{cp02}
%\ece
%\end{eqnarray}
\renewcommand{\member}{\mbox{\it member}}
From the conjunction
$\new(Ls_1,[B|Ls_2],Ls_3)\wedge \minlist(A,Ls_3)$
in the body of Clause
(\ref{cp01}) we can obtain the following consequences: 
a) $B\in Ls_3$ because
%We observe that\\ 
%\noindent %$(\member(X,Ls1) \vee \member(X,Ls2)) \wedge \shuffle(
%Ls_1,Ls_2,Ls_3) \implies \member(X,Ls_3)$
\[(X\in Ls_1 \vee X\in Ls_2)\wedge \shuffle(Ls_1,Ls_2,Ls_3) \implies
X\in Ls_3;\]
 b) $A\le X,\ \forall X . X\in Ls_3$ 
by the mathematical 
definition of \minlist; c) from a) and b), 
$A\le B$. Thus, we can add the inequality $A\le B$ to the body of 
Clause (\ref{cp01}). By a similar argument, we can add the 
inequality $B\le A$ to the body of Clause (\ref{cp02}).

\begin{eqnarray}
& & \new([A\br Ls_1]\co [B\br Ls_2]\co [A\br Ls_3]) \la 
        A\le B\noln
        \new(Ls_1\co [B\br Ls_2]\co Ls_3)\wedge  \minlist(A,Ls_3)\label{prob01}\\ 
& & \new([A\br Ls_1]\co [B\br Ls_2]\co [B\br Ls_3]) \la
        B\le A\noln
	\new([A\br Ls_1] \co Ls_2\co Ls_3)\wedge \minlist(B,Ls_3)\label{prob02} %\noln
\end{eqnarray}
The subgoal $\minlist(A,Ls_3)$ in Clause 
(\ref{prob01}) is unnecessary for the following reasons:
First, we observe that the the first two arguments 
of \new\ are already ordered (see Clause (\ref{alreadyordered})). Second, 
from Clause (\ref{cp01}), we have a) $A\le X,\ \forall X.X\in Ls_1$;
b) $B\le Y,\ \forall Y.Y\in Ls_2$; c) $A\le B\implies A\le Y,\ \forall Y.Y\in Ls_2$;
d) from a) and c), we have $A\le Z,\ \forall Z.Z \in Ls_3$ (because $Ls_3$ is 
conformed by elements belonging to $Ls_1$ and $Ls_2$). 
Hence, we can get rid of $\minlist(A,Ls_3)$ without losing correctness.
A similar argument works for Clause (\ref{prob02}).

Finally, we arrive at the following program:

\begin{program}[Mergesort algorithm] \label{sorting:merge}
\begin{eqnarray}
& & \msort(\nil \co \nil ) \la \\
& & \msort([A]\co [A])\la \\
& & \msort([A,B\br Ls_1]\co Ls_6) \la %\newln
        \divid([A,B\br Ls_1] \co Ls_2\co Ls_3) \noln
        \msort(Ls_2\co Ls_4)\wedge 
        \msort(Ls_3\co Ls_5) \noln
        \new(Ls_4\co Ls_5\co Ls_6)\\[\vsp]
& & \mbox{plus definitions of \divid/3 and \new/3 (without \minlist)} %\nonumber
%& & \divid(\nil \co \nil \co \nil ) \la \\
%& & \divid([A]\co \nil \co [A])\la \\
%& & \divid([A,B\br Ls_1]\co [A\br Ls_2]\co [B\br Ls_3])\la                
%\divid(Ls_1\co Ls_2 \co Ls_3)%\\[\vsp]
%& & \new(\nil \co L\co L) \la \ord1(L) \label{sorting:d1}\\
%& & \new(L\co \nil \co L) \la \ord1(L) \label{sorting:d2}\\
%& & \new([A\br Ls_1]\co [B\br Ls_2]\co [A\br Ls_3]) \la A\le B \comma
%	\new(Ls_1\co [B\br Ls_2]\co Ls_3) \\
%& & \new([A\br Ls_1]\co [B\br Ls_2]\co [B\br Ls_3]) \la B< A \comma
%	      \new([A\br Ls_1]\co Ls_2 \co Ls_3)
\end{eqnarray}
\end{program}
This program has the essential structure of the mergesort algorithm, 
and has an $O(n\log n)$ time complexity order.

\section{The  quicksort algorithm}

To derive the quicksort algorithm, we need to see again 
the Prog.~\ref{sorting:merge}. There,
our choice of the  $\divid(L,Ls_1,Ls_2)$ predicate was
motivated by  $L=Ls_1\cup Ls_2$, without any major constraint.
 Now we select another
definition, arguably more discriminant that the first one:
\begin{eqnarray}
& &  \selection([A\br Ls]\co Ls_1\co [A\br Ls_2]) \la \newln
   [A\br Ls]= Ls_1\dplus
  [A\br Ls_2]\comma \noln Ls_1\rex A \comma A \rex Ls_2  \label{sorting:sel}
%  & \nonumber\\
%& & \mbox{ where } Ls_1 = \{B\in L | B<A\} \nonumber \\ 
%& & \mbox{ and } Ls_2=\{B\in L | B>A\}
\end{eqnarray}

If $X=[A|Ls]$, $\selection([A|Ls]\co Ls_1\co [A|Ls_2])$ allows that
 $X=\set{Ls_1}\cup \set{[A|Ls_2]}$.
Our new choice allows us to eliminate the following (useless) clauses:
\be
& & \new(Ls\co \nil \co Ls) \la \\
& & \new([A\br Ls_1]\co [B\br Ls_2]\co [B\br Ls_3]) \la %\newln 
B<A\comma
\new([A\br Ls_1]\co Ls_2\co Ls_3) 
 \ee
Now, we  rename \new\ to $\append'$:
\be
& & \append'(\nil \co Ls\co Ls) \la \\
& & \append'([A\br Ls_1]\co [B\br Ls_2]\co [A\br Ls_3]) \la \newln
        B<A \wedge
        \append'(Ls_1\co [B\br Ls_2]\co Ls_3) 
\ee
Our next step is to eliminate the (unnecessary) comparison
 $B<A$.
%\footnote{How? And how to eliminate $\ord1(L)$ in
% clauses~\eqref{sorting:d1} and~\eqref{sorting:d2}?
% Currently, I am working these problems.}. 

To implement our new predicate we define the following
clauses:
$L$ is divide into two lists,  according
to~\eeqref{sorting:sel}:
\be
& & \selection(A\co\nil \co\nil \co\nil ) \la \\
& & \selection(A\co [B\br L]\co [B\br Ls_1]\co Ls_2) \la \newln
       B<A\comma \selection(A\co L\co Ls_1\co Ls_2)\\
& & \selection(A\co [B\br L]\co Ls_1\co [B\br Ls_2]) \la \newln
       B>A \comma \selection(A\co L\co Ls_1\co Ls_2)
\ee 
Therefore, we have derived the quicksort algorithm.
%{\raggedleft$\Box$}

%% \section{Exchanging algorithm}

%% By directing equation \cite{ChengPE95} we can express succinctly 
%% a sorting algorithm based on \emph{exchanging elements}.
%% The following algorithm, resembling the bubble-sort algorithm 
%% (it \emph{is} the bubble-sort algorithms over lists), has 
%% resisted an analogous treatment to the previous ones:
%% \begin{verbatim}
%% mysort(Ls,Ms) :- eq1(Ls,Ms).

%% eq1(Ls,Ms) :- eq2(Ls,Zs), eq1(Zs,Ms).
%% eq1(Ls,Ls).

%% eq2(Ls,Ms) :-  
%%     \+ord(Ls),
%%     exchangeList(Ls,Ms).

%% ord([]).
%% ord([A]).
%% ord([A,B|Ls]) :- A=<B, ord([B|Ls]).

%% exchangeList([],[]).
%% exchangeList([A],[A]).
%% exchangeList([A,B|Ls],[A|Ms]) :- A<B, exchangeList([B|Ls],Ms).
%% exchangeList([A,B|Ls],[B|Ms]) :- B=<A, exchangeList([A|Ls],Ms).
%% \end{verbatim}

\section{Comparison with similar work}

%Goal introduction in logic program transformation is often seen 
%as uncontrolled technique for program derivation \cite{}.

The ideas of deriving sorting algorithms have been
carried out through ma\-the\-ma\-tical-oriented developments \cite{Darlington:sssa},
logic program \cite{ClarkDarlington78} derivation synthesis,
and functional program transformation \cite{Partsch91}.
In work \cite{ClarkDarlington78}, some justifications were absent and are given
within our approach. 

%An identification of transformation rules 
%A classification of sorting algorithms as a consequence of derivations
% ordering properties and sorting 
%algorithms were derived within logic programming, but some points were
%left pendent or without concrete logic programs.
%
%It is worth noticing that many imperative sorting algorithms are based
%on doing a heavy use of destructive assigment and arrays, so that
%lists are not suitable to deal with them. 
%In \cite{Dromey87} there is
%a valuable set of sorting algorithms derived from the naive sorting
%algorithm within an imperative setting.
%

In~\cite{Lau:nscsal} and \cite{LP:sfrsp} the program 
synthesis proceeds as
follows: For each derivation, the user gives a 
general scheme in clausal
form, following a catalog of possible recursion patterns.
However, neither dependency with respect to supporting definitions 
nor correctness are shown;
also, in~\cite{LP:sfrsp} there is not any permutation
algorithm, and the base cases are added manually.

In~\cite{Moller:acgsa} there is a derivation of the
mergesort algorithm by using formal languages, 
but without following a computing paradigm and without any
commitment with a specific implementation. We %have  
adapted from \cite{Moller:acgsa}
the definition of \shuffle\ to logic programming. 
%Newly, however, the work of
%M\"oller, unlike ours, does not have any compromise with a specific
%implementation of his final programs.

%In~, Partsch
%derives some sorting algorithms, but he does his derivations using
%program synthesis instead program transformation (he does not define

%any permutation algorithm).

\section{Conclusions}

In this work we have presented the novel concept of abductive folding
as a mechanism to overcome some limitations of the unfold/fold method.
Abductive folding is carried out through two consecutive steps: an
application of the subgoal introduction rule and an application of the
traditional folding rule.
We have achieved some characterizations to identify suitable atoms to
be added to the body of clauses. 
Some sorting algorithms were derived: A
sorting algorithm devised by Tamaki and Sato, 
the selection sort algorithm, the insertion sort algorithm,
the mergesort algorithm and the quicksort algorithm.
% Some difficulties in the derivation of
%these algorithms are: the justification of introduction of atoms
%within the body of clauses, the fidelity in following some specific
%permutation and order algorithms to reach the structure of known
%algorithms, and the identification of properties to allow the
%replacement of subgoals when it is required.%
%This work has presented 
Some of the derivations of these sorting algorithms were carried out through
LPT with an occasional and complementary ALP support.
ALP has been applied for justifying the selection and the
introduction of atoms within the body of clauses, adapting the methods
of reasoning belonging to ALP to LPT.
This is the part of including non-declarative
heuristic and operational control to explain or refine a 
purely declarative problem description via a logical model \cite{alp1993}.

Tamaki and Sato in \cite{TS:uftlp} gave a short % and convincent
derivation of an $O(n^3)$ sorting algorithm from a naive sorting
algorithm by a technique of \emph{introducing} an atom in the body of
a clause.  However, Tamaki and Sato's technique has not been
recognized as a general and useful technique 
within LPT methodologies. Our
contribution is to argue that Tamaki and Sato's technique
can be identified as a valuable instance of the ALP approach 
for complementing LPT techniques.

%Because constraints in the way of inequalities appear within a sequential
%flow of computation, sometimes attaching a \emph{mode}
%\cite{Apt:97} will be necessary to guarantee the suitable
%instantiation and to overcome faulty circumstances related to an
%insufficient instantiation of arguments of the predicate $<$ and
%derived. %Through a consistent use of indexes we have mimicked...(IMPORTANTE)

Further research is required to mechanize (at least partially) the
selection and introduction of subgoals. The abductive task,
appa\-rent\-ly, depends on the presentation theory and some properties
explicitly formulated; these properties should involve the predicates
occurring within the logic programs at hand.  In LPT, the problem of
finding abductive explanations for some ``dead ends'' of derivations
seems promising, powerful and mechanically plausible.

%As future work, we will deal with  the insertion algorithm, the exchanging
%algorithm and the quicksort algorithm within our formalism.

%\section*{Acknowledgments}
%I happily thank the facilities given by IIMAS, UNAM.
%I would like to thank to David Rosenblueth by his
%formulation of $\shuffle$ in clausal form.  
%\bibliography{/home/m/bib/bm_bib}

%% \section*{Acknowledgments}
%% We gratefully acknowledge the facilities provided by 
%% \emph{Universidad 
%% Tecnol\'ogica de la Mixteca} and
%%  \emph{Universidad Nacional Aut\'onoma de M\'exico}.
%% Manuel Hern\'andez thanks \emph{Conacyt} for its valuable support.
%% %was supported by a grant provided 
%% %by \emph{Conacyt}.

\bibliography{appmath}
\end{document}